\documentclass{aa}

\usepackage{graphicx}
\usepackage{txfonts}
\usepackage{hyperref}
\usepackage{multirow,bigdelim}
\usepackage{bm}
\usepackage{float}
\usepackage{placeins}

\usepackage{soul}

\newcommand{\vsin}{\hbox{$v \sin i$}} 
\newcommand{\kms}{\hbox{km\,s$^{-1}$}} 

\begin{document}

\title{The crucial role of surface magnetic fields for stellar dynamos: \\
$\epsilon$ Eridani, 61 Cygni A, and the Sun} 
\author{S.~V.~Jeffers\inst{1}
\and R.~H.~Cameron \inst{1} 
\and S.~C.~Marsden\inst{2} 
\and S.~Boro Saikia\inst{3} 
\and C.~P.~Folsom\inst{4}
\and M.~M.~Jardine\inst {5} 
\and J.~Morin\inst{6} 
\and P.~Petit\inst{7} 
\and V.~See\inst{8,9} 
\and A.~A.~Vidotto\inst{10,11} 
\and U.~Wolter\inst{12} 
\and M.~Mittag\inst{12}}


\institute{
Max-Planck-Institut f\"ur Sonnensystemforschung, Justus-von-Liebig-weg 3, 37077 G{\"o}ttingen, Germany
\and  Centre for Astrophysics, University of Southern Queensland, Toowoomba, Queensland 4350, Australia
\and University of Vienna, Department of Astrophysics, T\"urkenschanzstrasse 17, 1180 Vienna, Austria
\and Tartu Observatory, University of Tartu,Observatooriumi 1, T\~oravere, 61602 Tartumaa, Estonia 
\and SUPA, University of St Andrews, School of Physics and Astronomy, North Haugh, St Andrews KY16 9SS, UK
\and Laboratoire Univers et Particules de Montpellier, Centre national de la recherche scientifique (CNRS), Université de Montpellier, Place Eugène Bataillon, F-34095 Montpellier, France 
\and Institut de Recherche en Astrophysique et Plan\'etologie, Universit\'e de Toulouse, 14 ave Edouard Belin, F-31400 Toulouse, France  
\and European Space Agency (ESA), European Space Research and Technology Centre (ESTEC), Keplerlaan 1, 2201 AZ Noordwijk, The Netherlands
\and University of Exeter, Department of Physics \& Astronomy, Stocker Road, Exeter, Devon, EX4 4QL, UK
\and  Trinity College Dublin, Dublin 2, Ireland
\and Leiden Observatory, Leiden University, PO Box 9513, 2300RA, Leiden, The Netherlands
\and Hamburger Sternwarte, Universit\"at Hamburg, Gojenbergsweg 112,D-21029 Hamburg, Germany
}
\date{Received <-->; accepted <-->}

\abstract{
Cool main-sequence stars, such as the Sun, have magnetic fields which are generated by an internal dynamo mechanism.  In the Sun, the dynamo mechanism produces a balance between the amounts of magnetic flux generated and lost over the Sun's 11-year activity cycle and it is visible in the Sun's different atmospheric layers using multi-wavelength observations.  We used the same observational diagnostics, spanning several decades, to probe the emergence of magnetic flux on the two close by, active- and low-mass K dwarfs: 61~Cygni~A and $\epsilon$~Eridani.  Our results show that 61~Cygni~A follows the Solar dynamo with a regular cycle at all wavelengths, while $\epsilon$~Eridani represents a more extreme level of the Solar dynamo, while also showing strong Solar-like characteristics. For the first time we show magnetic butterfly diagrams for stars other than the Sun.  For the two K stars and the Sun, the rate at which the toroidal field is generated from surface poloidal field is similar to the rate at which toroidal flux is lost through flux emergence. This suggests that the surface field plays a crucial role in the dynamos of all three stars.   
Finally, for $\epsilon$~Eridani, we show that the two chromospheric cycle periods, of $\sim$3 and $\sim$13 years, correspond to two superimposed magnetic cycles.}


\maketitle 

\section{Introduction}

Magnetic activity is ubiquitous on the Sun and other solar-type stars. The Sun's varying magnetic activity is well established to be generated by an internal dynamo with a cyclic periodicity of approximately 11 years as revealed by multi-wavelength synoptic observations that probe the different layers in the Sun's atmosphere.  These observations range from broadband visible wavelength observations that monitor the Sun's activity in the form of dark sunspots \citep{1844AN.....21..233S},  chromospheric emission in several atomic lines \citep{2017ApJ...835...25E}, to coronal X-ray
emission \citep{1994SoPh..154..275G}. 
The magnetic nature of the Sun's activity patterns has been confirmed using polarimetric observations \citep{1908ApJ....28..315H}, and regular synoptic observations during the last 40 years have revealed the spatio-temporal structure of the Sun's large-scale magnetic field
\citep{Cameron2018A&A...609A..56C}.  The combination of these multi-wavelength diagnostics shows that the normal cyclic mode of the solar dynamo, as monitored with observations of sunspots over the last 400 years approximately,
can be explained by the Babcock-Leighton dynamo model where magnetic flux is generated and lost over the Sun's 11-year cycle  \citep{1961ApJ...133..572B, 1969ApJ...156....1L}. 

The challenge for Solar dynamo models is to explain both the high levels of activity that we observe on other stars, in the form of higher starspot coverage and coronal X-ray
emission \citep{Strassmeier2009A&ARv..17..251S,Guedel2009A&ARv..17..309G}, and periods of grand minima in Solar activity, such as the Maunder minimum in the 1600s.  Our motivation is to understand how similar are solar and stellar dynamos.

\FloatBarrier
\begin{table*}[]
\caption[]{{\bf{Fundamental stellar parameters of $\epsilon$~Eridani, 61~Cygni~A and the Sun.}}}
\centering
\begin{tabular}{lccccc}
\hline
\hline
Star                      &  $\epsilon$~Eridani$^{a,b}$       &  61~Cygni~A$^{c,d}$ & Sun $^e$ \\
\hline
HD                      &  22040                 &  201091           & -- \\
Spectral Type             & K\,2\,V                   & K\,5\,V           & G\,2\,V    \\
T$_{\rm eff}$ (K)         & $5146\pm 31$           &  $4545\pm 40$  & $5780\pm 10$      \\
Mass (M$_\odot$)          & $0.82\pm0.03$ &  $0.69\pm0.03 $       & 1.00    \\
Age: V\&F$^f$ (Gyr)       & $2.6^{+5.8}_{-2.4}$    &  --            & --    \\
Age: Chromo$^g$ (Gyr)     & $0.49$                &  $1.33$     & --      \\
Age: other (Gyr)          & {0.44}          &  {6.00 $\pm$ 1.00}  & $4.57 \pm 0.11 $\\
\vsin\ (\kms)             & $2.40 \pm 0.4$          &  $0.94\pm 1.0$         & $1.60 \pm 0.4$    \\
$P_{\rm rot}$ (d)& $11.40$                   &  $35.00$   &  $24.47$          \\
Rossby No    &        0.44                   &  1.41       & 2.18        \\
Log(R$^{\prime}_{\rm HK}$)& --4.40 to --4.50         &  --4.51 to --4.67  & --4.85 to --5.00 $^i$   \\
S-index cycle $^{j,k}$(yrs)   & $2.95\pm0.03$, $12.70\pm0.30$ & $7.30\pm0.1$ & $11.04\pm1.20$ \\
Log($L_{\rm x}/L_{bol}$) &  --4.80 & --5.58 & --6.24   \\
\hline
\hline
\end{tabular}
\\
\tablebib{Listed are the effective temperature (T$_{\rm eff}$), mass, age and projected rotational velocity (\vsin) of the stars. $a$,\cite{Valenti2005ApJS..159..141V}; $b$,\cite{Barnes2007ApJ...669.1167B}; $c$,\cite{Affer2005A&A...433..647A}; $d$,\cite{Kervella2008A&A...491..855K}; $e$,\cite{Wright2011ApJ...743...48W}; $f$,\cite{Valenti2005ApJS..159..141V}; $g$, \cite{Marsden2014MNRAS.444.3517M}; $h$, \cite{Metcalfe2013ApJ...763L..26M}; $i$, \cite{Lockwood2007ApJS..171..260L}; $j$, \cite{Baliunas1995ApJ...438..269B}; $k$, \cite{Bonanno2002A&A...390.1115B}.  The convective turnover time, used to compute the Rossby number, and X-ray luminosity (Log($L_{\rm x}/L_{bol}$)) are taken from  \cite{Wright2011ApJ...743...48W}.}
\label{tab:param-stars}
\end{table*}

The internal structure of other G, K and early-M dwarfs is similar to that of the Sun with a radiative core and an outer convective zone.   For the case of the Sun,  the convective zone is a key component of how its dynamo operates.  While G, K and M dwarfs exhibit comparable signatures of magnetic activity as the Sun, the ranges of activity levels are much more extreme -- ranging from stars showing almost no signs of activity to stars with activity levels that are orders of magnitude greater than the Sun. 
By observing other stars, our aim is to understand if the solar-dynamo model can also generate more extreme levels of magnetic activity and if we can fill the observational gap in terms of the grand minima states of the Sun.\\

In this paper we analyse the large-scale magnetic field geometry, and the chromospheric S-index of the three stars, $\epsilon$~Eridani, 61~Cygni~A, and the Sun.
All three stars exhibit activity cycles with $\epsilon$~Eridani having two chromospheric activity cycles \citep{Metcalfe2013ApJ...763L..26M}. Further properties of these stars are given in Table~\ref{tab:param-stars}.
We show in Section~3 that the S-index cycles are related to cycles in the net axisymmetric component of the surface toroidal flux. On the Sun this toroidal flux is due to flux emergence. Assuming that the surface toroidal field on all three stars corresponds to flux emergence, in Section~4 we show that if the differential rotation of $\epsilon$~Eridani and 61~Cygni~A are comparable to that of the Sun, then the amount of net axisymmetric toroidal flux generated in the convection zone associated with the surface poloidal field is similar to the amount of toroidal flux that is lost through the surface. This suggests that the surface poloidal field plays a crucial role not only in the Sun's dynamo, but also in the dynamos of $\epsilon$~Eridani and 61~Cygni~A.  

\section{Magnetic field maps}

\subsection{The Sun}
\begin{figure}
\begin{center}
\includegraphics[scale=0.48]{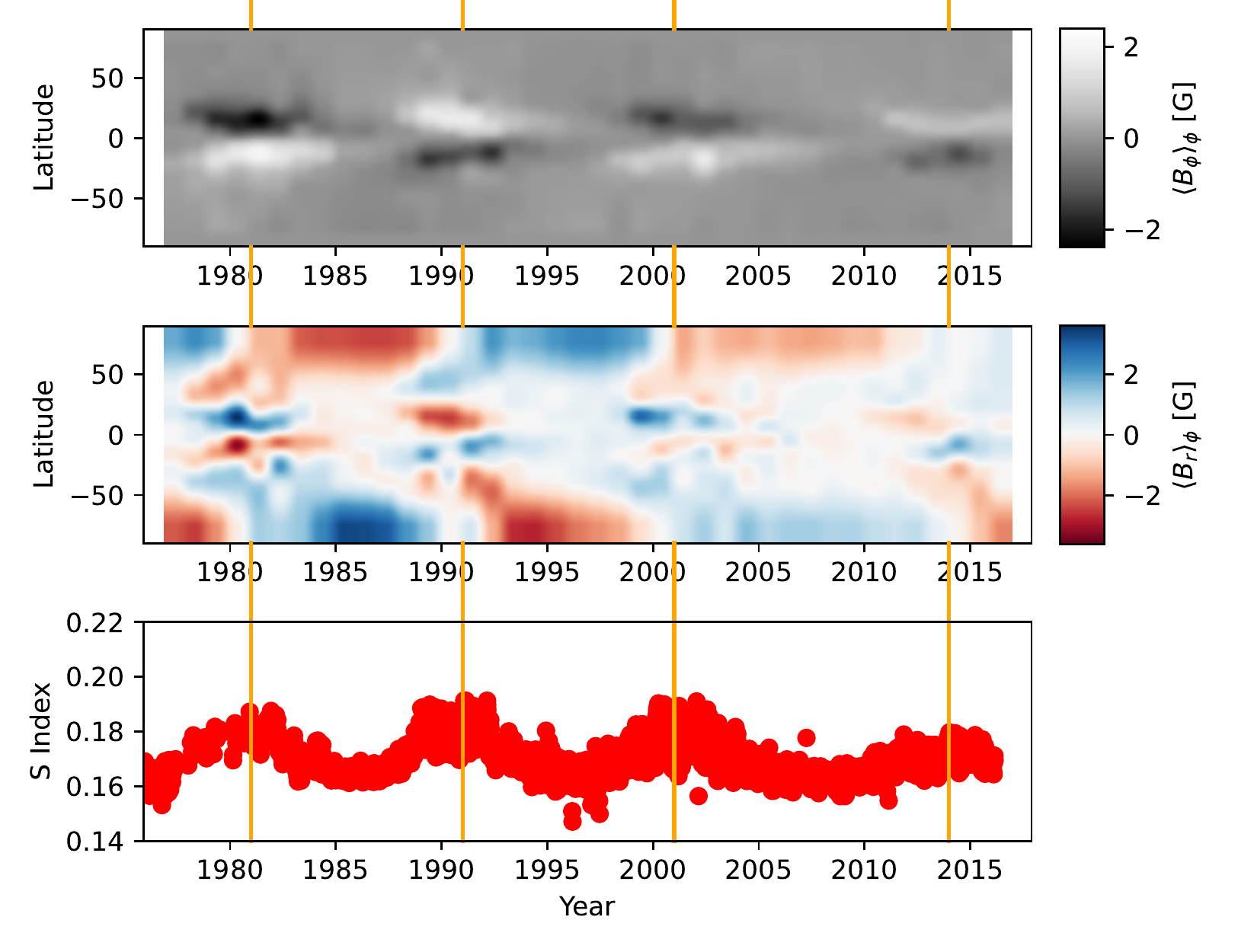}
\includegraphics[scale=0.48]{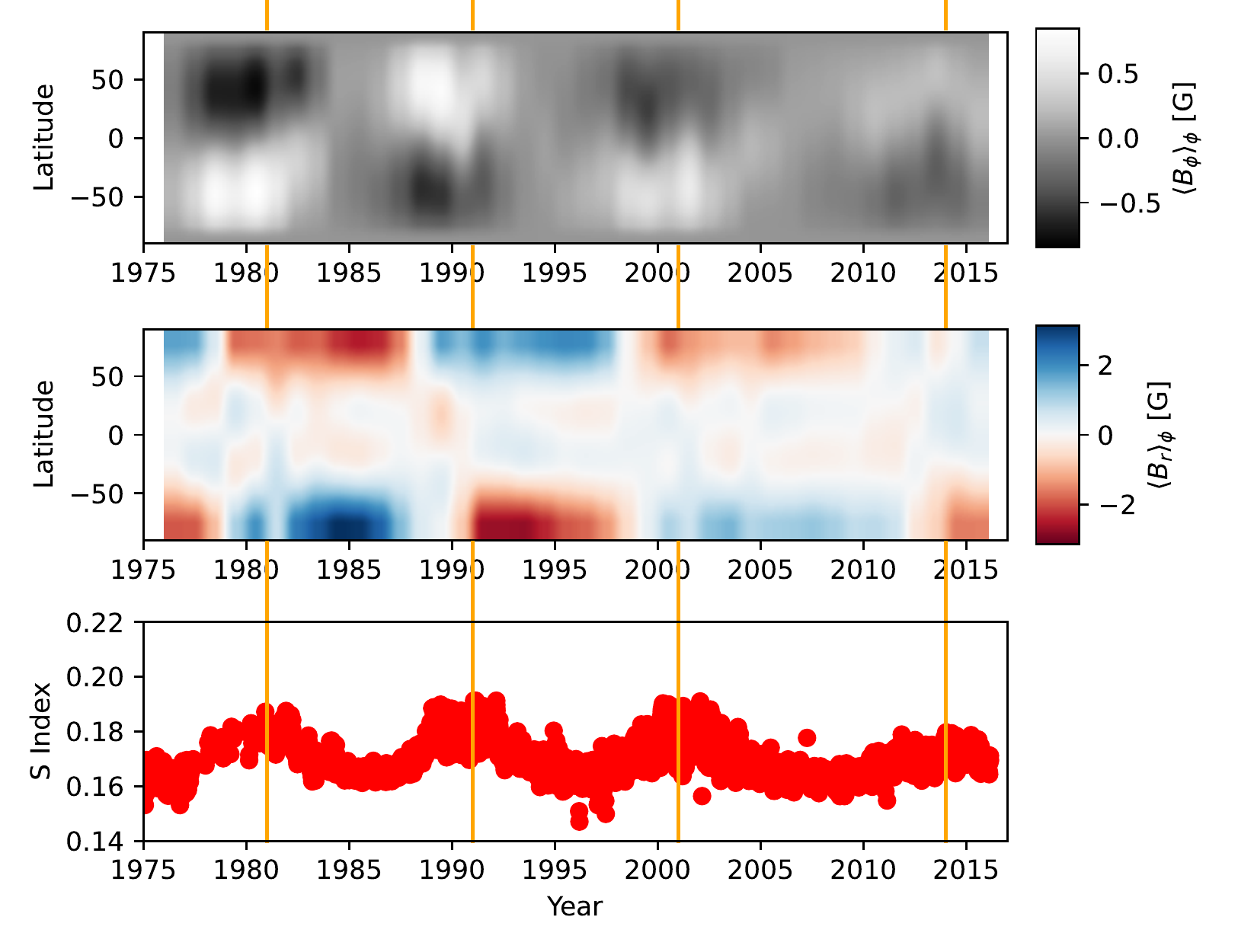}\\
\caption{Solar butterfly diagram from observations secured with the Wilcox Solar Observatory. The top three rows show the following:
$\langle B_\phi \rangle_\phi$ , 
with positive polarity in white and negative polarity in black (top panel);  $\langle B_r \rangle_\phi$ with positive polarity in red and negative polarity in blue (second panel);  and the chromospheric S-index values, where the yellow vertical lines indicate the S-index activity maximum (third panel).  The lower three panels are the same, but with the magnetic maps reduced to a lower resolution matching that achievable for 61~Cygni~A and $\epsilon$~Eridani.    
  \label{fig:Butterfly-Sun}
  }
\end{center}
\end{figure}

The Sun's line-of-sight magnetic field has been observed daily with the Wilcox Solar Observatory (WSO) since the mid-1970s covering more than four solar activity cycles. 
To determine the axisymmetric component of the radial and azimuthal fields, we exploit the fact that near the limb the line-of-sight component of the magnetic field is due to the azimuthal field, whereas near the meridian the line-of-sight magnetic field corresponds to a mixture of the radial and theta components of the field. 
This is the same basic property which is exploited by the tomographic technique of Zeeman-Doppler imaging (ZDI) \citep[][and references therein]{Donati2006MNRAS.370..629D,Folsom2018MNRAS.474.4956F}.  For the Sun, we followed the methods set out in
\cite{1979SoPh...61..233D} and \cite{Cameron2018A&A...609A..56C}. 
We first integrated full-disk magnetograms of the line-of-sight component of the field for 1 year. This largely averaged out the signal from  the non-axisymmetric components of the field. For each year, we obtained the full-disk, time-averaged  line-of-sight component of the magnetic field $B_{\mathrm{los}}$ as a function of latitude and longitude relative to the central meridian. 
We decomposed the line-of-sight component of the field into components which are symmetric and anti-symmetric in longitude relative to the central meridian. 
The antisymmetric component, $B_{\mathrm{los}}^A$, is related to the longitudinally 
averaged azimuthal field $\langle B_\phi(\theta)\rangle_\phi$ (where
$\langle ... \rangle_\phi$ indicates the azimuthal average), 
by $B_{\mathrm{los}}^A(\theta,\phi)
=\langle B_\phi(\theta)\rangle_\phi \sin{(\phi-\phi')},
$ where  $\theta$ is the latitude and
$\phi-\phi'$ is the longitude relative to the central meridian. 
We determined $\langle B_\phi(\theta)\rangle_\phi$ from the observations by fitting to the above relation. Related techniques were used by \cite{1994SoPh..153..131S,2010ASPC..428..109L,2005ApJ...620L.123U}.
Figure~\ref{fig:Butterfly-Sun} shows the magnetic butterfly diagrams for the Sun at the resolution of the WSO observations and at the reduced resolution when only spherical harmonics up to $\ell \le 3$ were kept.

\subsection{61~Cygni~A and $\epsilon$~Eridani}

The large-scale magnetic field geometry of magnetically active K dwarfs 61~Cygni~A and $\epsilon$~Eridani were reconstructed using ZDI from a time series of circularly polarised spectropolarimetric observations \citep[for more details see:][]{Jeffers2014A&A...569A..79J, BoroSaikia2016A&A...594A..29B,Jeffers2017MNRAS.471L..96J, Petit2021A&A...648A..55P}. The two K dwarfs have been regularly observed since 2005 with the NARVAL spectropolarimeter mounted on the Telescope Bernard Lyot at the Observatoire Midi-Pyrenees in France as part of the BCool long-term programme to understand the magnetic fields of solar-type stars \citep{Marsden2014MNRAS.444.3517M}. 61~Cygni~A and $\epsilon$~Eridani are the only the stars from the sample of G or K dwarfs that have both a complete S-index time series and with regular ZDI maps over their S-index cycles.\\

\begin{figure}
\begin{center}
\includegraphics[scale=0.48]{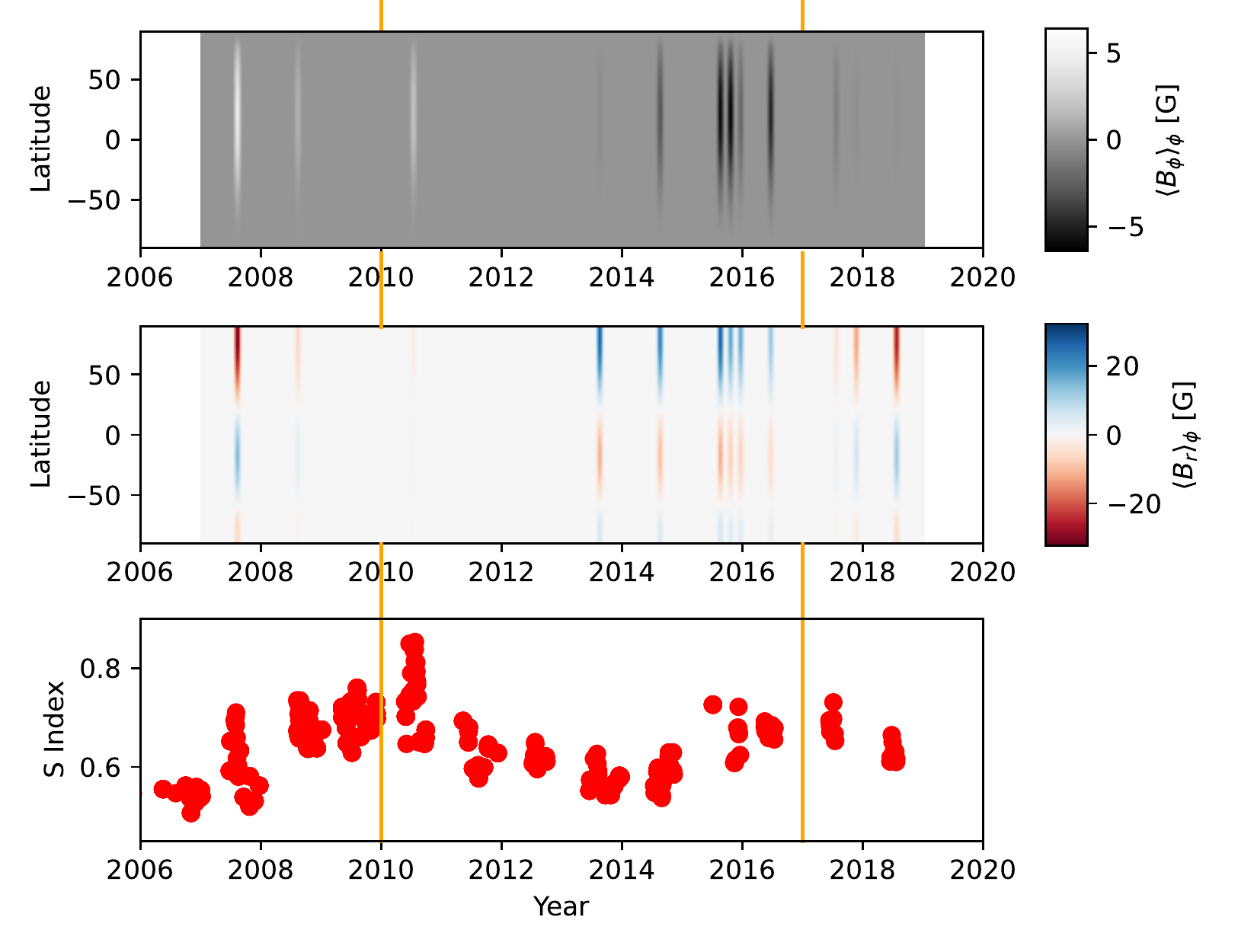}
\includegraphics[scale=0.48]{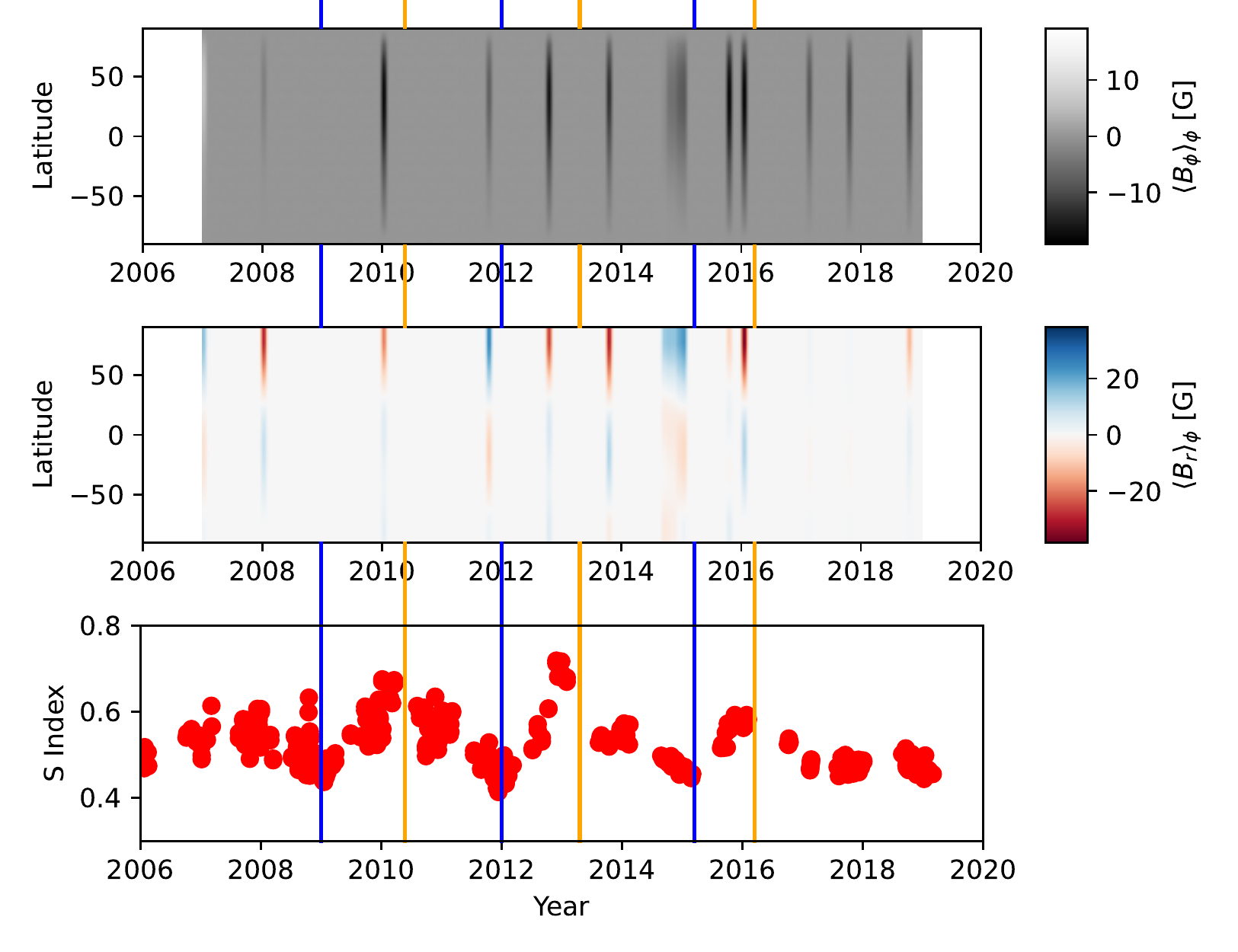}\\
\caption{Butterfly diagrams for 61~Cygni~A (upper three panels) and $\epsilon$~Eridani (lower three panels). The format is as described for Figure ~1.  The vertical blue lines indicate the S-index minima. 
\label{fig:Butterfly-stars}
}
\end{center}
\end{figure}

For this paper we only consider the axisymmetric components of the field which are well recovered by ZDI  \citep[previously shown by][]{Lehmann2021MNRAS.500.1243L}.  
The axisymmetric component of the magnetic field of 61~Cygni~A has been reconstructed for observations spaced at almost yearly epochs since 2005
and it is shown as butterfly diagrams in Figure~\ref{fig:Butterfly-stars}, for the radial and azimuthal components, along with the butterfly diagrams for $\epsilon$~Eridani.
Magnetic cycles are visible for both stars.

\section{Hemispheric surface toroidal flux versus chromospheric activity}
\begin{figure*}
\begin{center}
\includegraphics[scale=0.65]{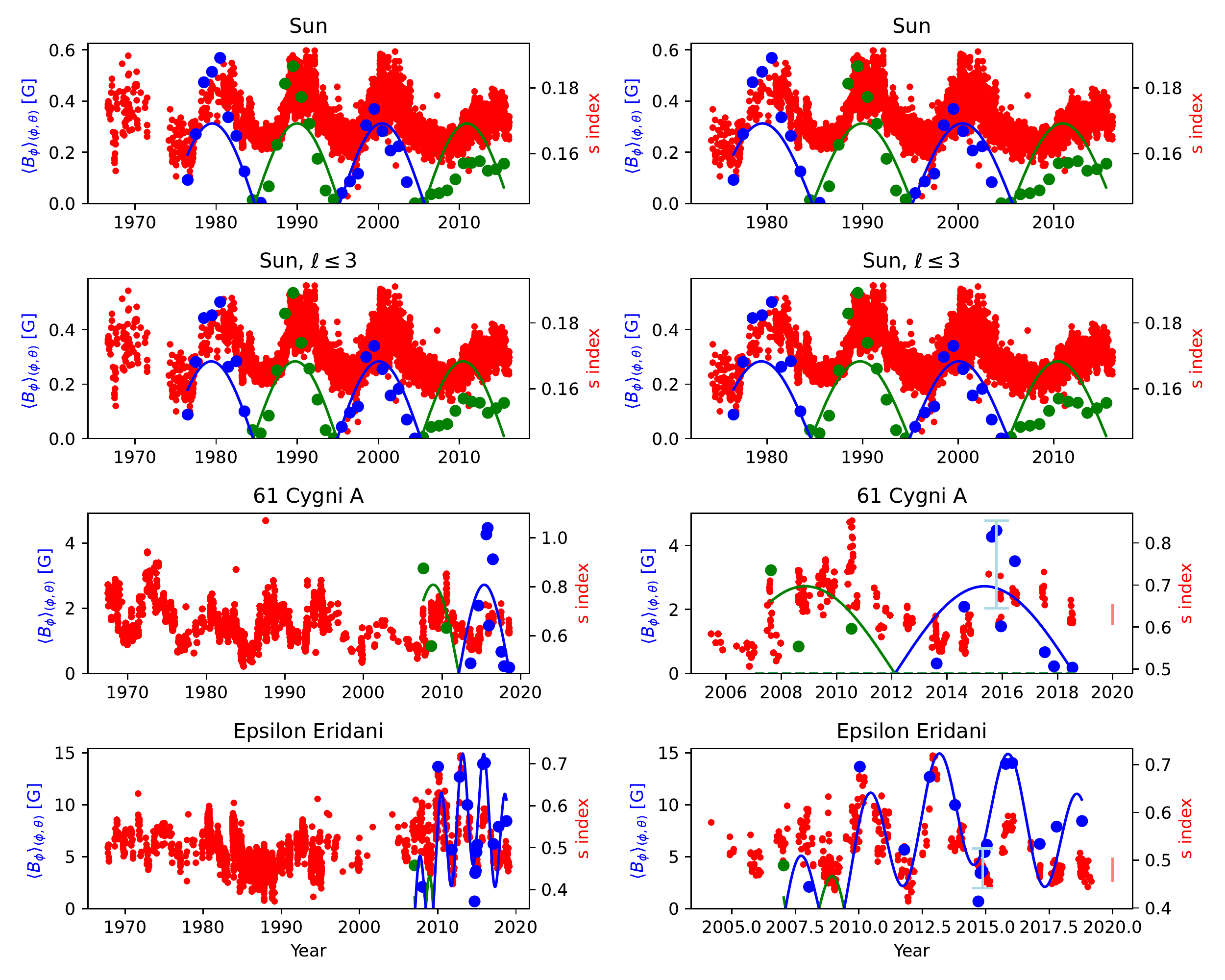}
\caption{Time evolution of flux emergence and chromospheric emission:  
The unsigned latitudinal and longitudinal average of the azimuthal surface magnetic field, $\langle B_\phi \rangle_{(\phi,\theta)}$, is shown in blue and green points indicating where the signed average is negative or positive, respectively. The solid curves are single sinusoidal fits to $\langle B_\phi \rangle_{(\phi,\theta)}$ in the case of the Sun and 61~Cygni~A, and a fit using a sum of two sinusoids for $\epsilon$~Eridani.
The light blue bars indicate the estimated error bars for $\langle B_\phi \rangle_{(\phi,\theta)}$, as described in the text.
The independently determined level of chromospheric activity as measured by the S-index is shown with red points and the vertical bar at year 2020 shows the S-index error bar.
The top row shows the Sun at high resolution, the second row shows the Sun at the equivalent resolution of the ZDI maps, and the third and fourth rows show the results for 61~Cygni~A and $\epsilon$~Eridani, respectively.
The left panel extends back to the S-index observations obtained in the 1960s while the right-hand panel zooms into the time period where there are also magnetic maps.    
\label{fig:S_vs_Bphi}
  }
\end{center}
\end{figure*}

In the solar case, magnetic flux emerging through the photosphere dominates the axisymmetric component
of the toroidal field observed at the surface $\langle B_\phi(\theta)\rangle_\phi$ \citep{Cameron2018A&A...609A..56C}.
The relationship between the S-index, a measure of chromospheric activity, and $\langle B_\phi \rangle_{(\phi,\theta)}$ (the latitudinally and longitudinally averaged surface azimuthal field) is shown in the top row of Figure~\ref{fig:S_vs_Bphi}. The Sun's chromospheric activity follows the emergence of flux with a delay that can be up to 10 months \citep{1994SoPh..150....1S}. Reducing the resolution of the magnetic field  observed on the Sun to that typically achievable for other stars was performed by projecting the observations onto the basis functions commonly used in ZDI inversions \cite{Donati2006MNRAS.370..629D,Folsom2018MNRAS.474.4956F} and only keeping the first three terms in the spherical harmonics expansion ($\ell \le 3$).  This reduction in resolution has little effect on the temporal evolution of the solar data as the correlation between $\langle B_\phi \rangle_{(\phi,\theta)}$ and the S-index still holds (see Figure~\ref{fig:S_vs_Bphi}).\\

Since stellar activity is also driven by magnetic fields, we investigated if the same correlations hold for 61~Cygni~A and $\epsilon$~Eridani.  
The S-index and the surface magnetic field  $\langle B_\phi \rangle_{(\phi,\theta)}$ are shown
in Figure~\ref{fig:S_vs_Bphi}, with
 $\langle B_\phi \rangle_{(\phi,\theta)}$ being
computed using only the first three terms ($\ell \le 3$) in the ZDI spherical harmonics expansion to allow for a direct comparison with the solar data.
This strongly suggests that $\langle B_\phi \rangle_{(\phi,\theta)}$ on all three stars is associated with flux emergence which subsequently drives chromospheric activity.  In agreement with this are the X-ray cycles, which are co-incident with the S-index cycle of 61~Cygni~A \citep{BoroSaikia2016A&A...594A..29B} and the short cycle for $\epsilon$~Eridani \citep{Coffaro2020A&A...636A..49C}. \\

For 61~Cygni~A, three ZDI maps cover the second half of 2015. We assumed that the differences in the fluxes of these three maps is due to noise, and from this we inferred the error bar shown in light blue (near year 2016) in Figure~\ref{fig:S_vs_Bphi}. For $\epsilon$~Eridani, five maps covering the second half of 2014 were used to estimate the random component of the error. We note that the error bars are for the unsigned field strength, which indicates that the cyclic behaviour of the unsigned flux is highly significant for all three stars.    

Two S-index activity cycles observed for $\epsilon$~Eridani, with periods of 2.95 and 12.7 years, are visible in the time series of $\langle B_\phi \rangle_{(\phi,\theta)}$. The shorter S-index period varies almost in phase with $\langle B_\phi \rangle_{(\phi,\theta)}$. This $\sim$3-year periodicity in $\langle B_\phi \rangle_{(\phi,\theta)}$ does not oscillate around zero but rather around a slowly varying level. The sign of $\langle B_\phi \rangle_{(\phi,\theta)}$ does not change polarity throughout most of the  $\sim$15 year time span covered by the observations.  The slowly varying component of $\langle B_\phi \rangle_{(\phi,\theta)}$ has an inferred period of about 28 years, corresponding roughly to twice the $\sim$13 year chromospheric activity period.  The longer periodicity is similar to both the cycles of the Sun and 61~Cygni~A. 
The short-period variability in $\epsilon$~Eridani has a period of about 3 years in both the chromospheric activity and $\langle B_\phi \rangle_{(\phi,\theta)}$. The emergence rate from the chromospheric data reflects the signed sum of the field associated with both the 2.95 and 12.7 year cycles. This is why the short-term chromospheric activity cycle and
the short magnetic cycle have the same period (rather than the magnetic period being twice the activity period).
A similar behaviour is seen on the Sun in the context of the quadrupole and dipole modes of the solar dynamo 
\citep{2018A&A...618A..89S}. 

\section{Toroidal flux budgets}

\begin{figure}
\begin{center}
\includegraphics[scale=0.5]{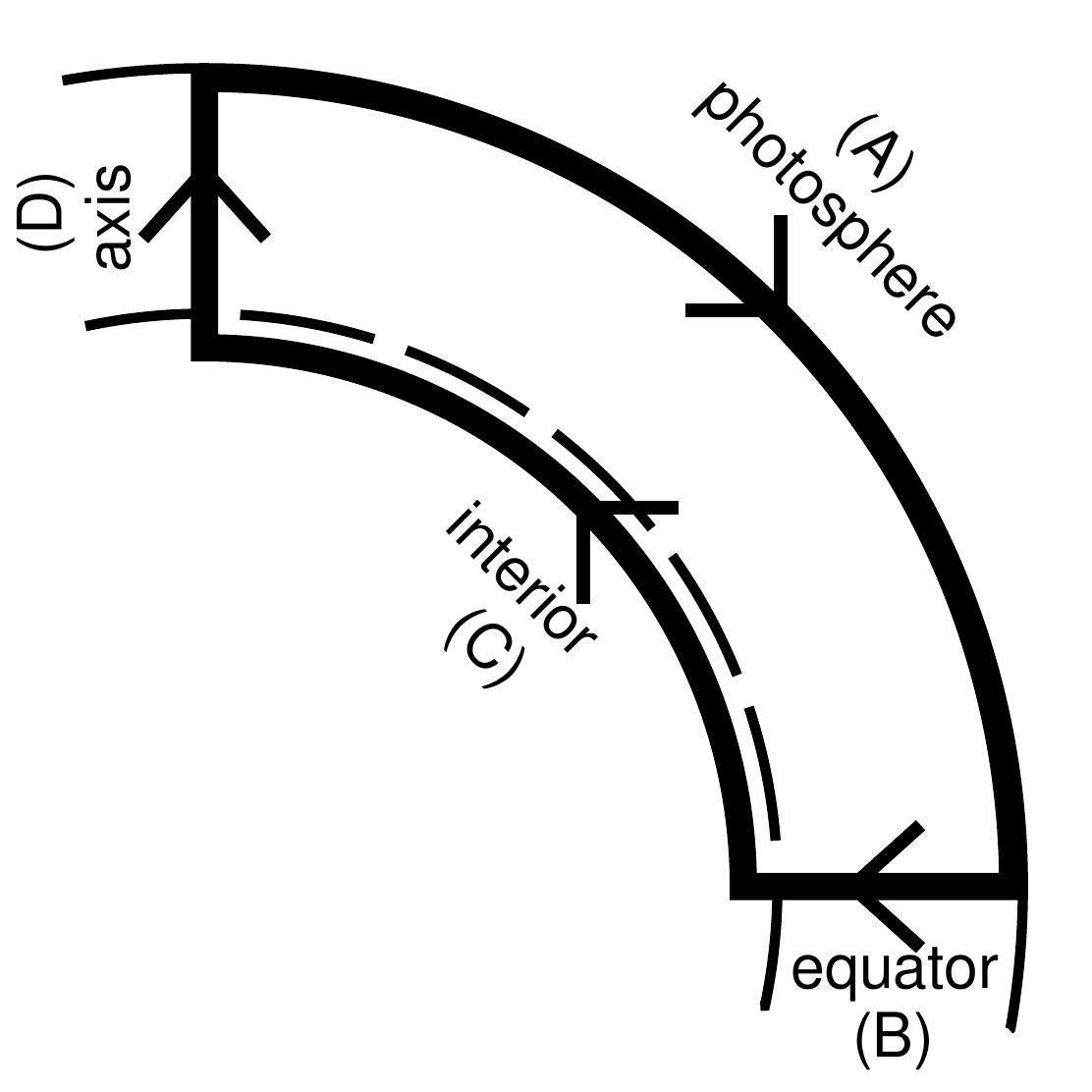}
\caption{Meridional cut through the convection zone in the northern hemisphere
of the Sun. The thin dashed curve indicates the bottom of the convection zone. The thin solid curve represents the photosphere.
The thick solid contour outlines the chosen surface. This choice follows \cite{2015Sci...347.1333C} and \cite{2020A&A...636A...7C}}.
\label{fig:contour}
\end{center}
\end{figure}

The evolution of the magnetic field, ${\bf B}$, inside the Sun and stars is governed by the induction equation:
\begin{eqnarray}
\frac{\partial{\bf{B}}}{\partial t}= \nabla \times({\bf U} \times {\bf B}+\eta \nabla \times {\bf B}),
\end{eqnarray}
where ${\bf U}$ is the velocity and $\eta$ is the microscopic diffusivity. Following \cite{2015Sci...347.1333C}, we chose to work in a coordinate system rotating at the same rate as the surface at the equator for each star. We applied Stokes theorem to a
meridional cut through the convection zone bounded by the contour  shown in Figure~\ref{fig:contour}. 
This yielded
\begin{eqnarray}
\frac{\partial{\int_A\langle {\bf{B}} \rangle_\phi \cdot \mathrm{d}{\bf A}}}{\partial t}= \int_{\delta A} \langle {\bf U} \times {\bf B}\rangle_\phi \cdot \mathrm{d}\boldsymbol{\ell}
\end{eqnarray}
where we dropped the term involving the microscopic diffusivity on the right-hand side because $\eta$ is tiny. We note that $\int_A\langle {\bf{B}} \rangle_\phi \cdot \mathrm{d}{\bf A}$ is the azimuthally averaged toroidal flux in the hemisphere.
We can further write ${\bf U}=\langle{\bf U}\rangle_\phi +{\bf U'}$ and ${\bf B}=\langle{\bf B}\rangle_\phi +{\bf B'}$ in terms of their axisymmetric means and non-axisymmetric components.
We then obtained
\begin{eqnarray}
\frac{\partial{\int_A\langle {\bf{B}} \rangle_\phi \cdot \mathrm{d}{\bf A}}}{\partial t}= \int_{\delta A}
\left(
\langle {\bf U}
\rangle_\phi \times \langle{\bf B}\rangle_\phi
+
\langle {\bf U'} \times {\bf B'}\rangle_\phi
\right)
\cdot
\mathrm{d}\boldsymbol{\ell}
\label{eq:stokes2}
\end{eqnarray}
noting that $\langle \langle U \rangle_\phi \times B'\rangle_\phi=
\langle U' \times \langle B \rangle_\phi  \rangle_\phi=0$.

Following \cite{2015Sci...347.1333C} and \cite{2020A&A...636A...7C}, we considered the contributions from the two terms on the right-hand side of Eqn~\ref{eq:stokes2} separately. Beginning with 
$\int_{\delta A} \langle {\bf U}
\rangle_\phi \times \langle{\bf B}\rangle_\phi \cdot \mathrm{d}\boldsymbol \ell$, the dynamo-related component of the integrand vanishes along segments (C) and (D)
shown in Figure \ref{fig:contour}  because segment (C) was chosen to be below the level where the cycle-related field penetrates, also the axisymmetric $\phi$ and $\theta$ 
components of the field and flow vanish along rotation axis (D).
Thus the non-negligible contributions can only come from the surface and from the line segment at the equator. 
The contribution from line segment (B) is small on the Sun because the radial differential rotation at the equator is small. Whether this term is also small on other stars depends on their radial differential rotation.
The contribution from the integral at phostosphere (E) is given by
$-\int_0^{\pi/2} \langle B_r \rangle (\Omega(\theta)-\Omega(\pi/2)) R_\star^2 \sin \theta \mathrm{d}\theta$, where $\Omega(\theta)$ is the surface rotation rate at colatitude $\theta$, and $\langle B_r \rangle$ is the azimuthally averaged surface radial field
\citep{2015Sci...347.1333C}, and it can be evaluated from observations of the surface radial magnetic field if the surface differential rotation is known.

The other half of the right-hand side of Eqn~\ref{eq:stokes2}
gives the contribution from small-scale motions, $\int_{\delta A} \langle {\bf U'} \times {\bf B'}\rangle_\phi \cdot\mathrm{d}\boldsymbol \ell$. The contributions to this term from the parts of $\delta A$ along (B) and (D) are expected to be diffusive in nature due to symmetry arguments. The contribution from (C) is expected to vanish because the cycle-related component of ${\bf{B}'}$ should vanish there. 
This leaves an integral along the photosphere. As is shown in \cite{2020A&A...636A...7C}, a straightforward implication of the empirical flux transport model is that this integral is determined by flux emergence events. 
We also note that the $\phi$-component of the axisymmetric magnetic field seen at the solar photosphere corresponds to flux emerging through the surface \citep{Cameron2018A&A...609A..56C}.
This allows us to write the integral as  
$-\int_0^{\pi/2} u_\mathrm{em} R \langle B_\phi\rangle_{(\phi,\theta)} \mathrm{d}\theta $ where $u_\mathrm{em}$ is the velocity in the radial direction associated with the magnetic field as it emerges through the photosphere. Thus the amount of flux loss associated with emergences is estimated by $\int_0^{\pi/2} u_\mathrm{em} R \langle B_\phi\rangle_{(\phi,\theta)} \mathrm{d}\theta $.

\begin{figure}
\begin{center}
\includegraphics[scale=0.75]{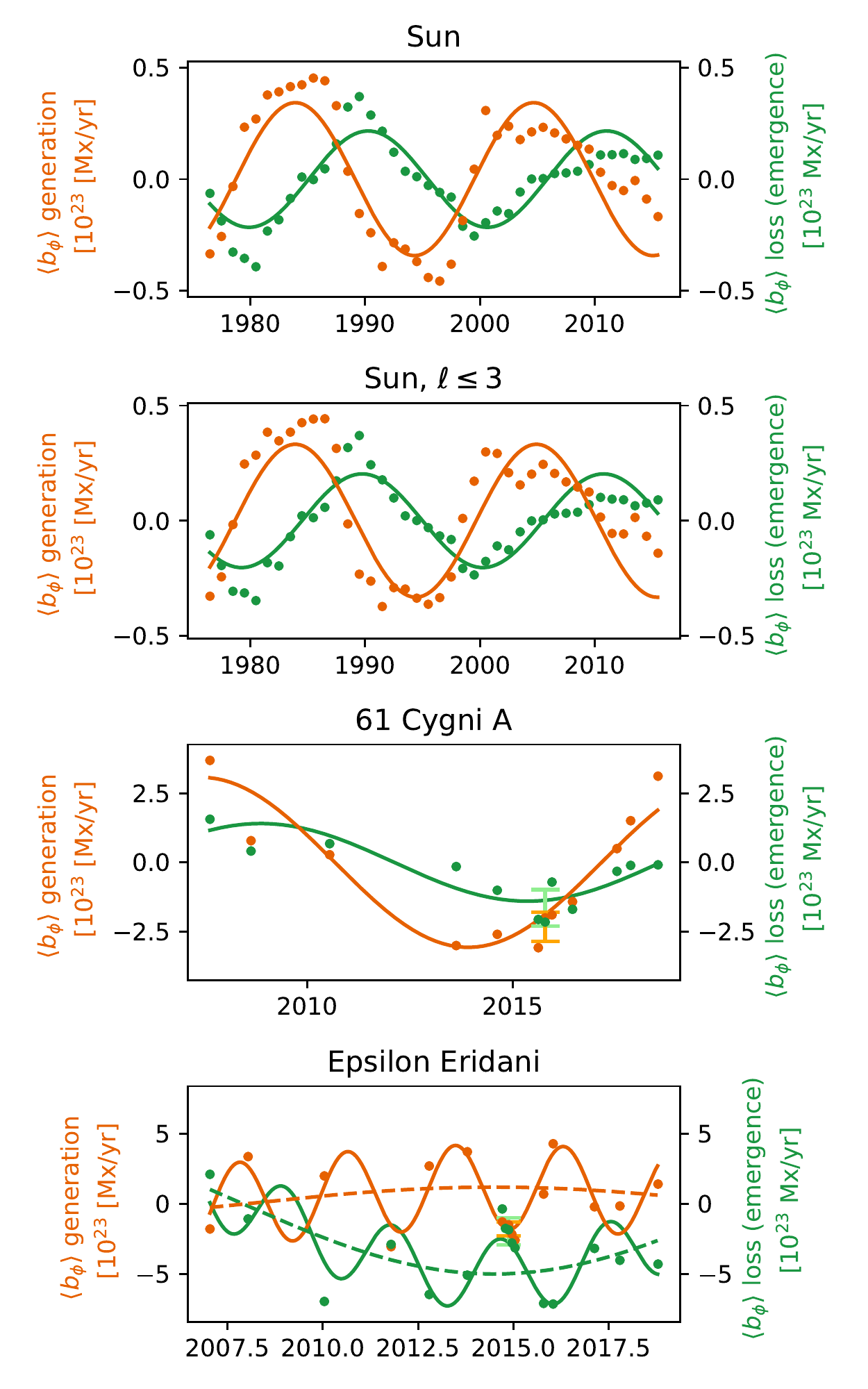}
\caption{Toroidal magnetic field generation and loss associated with surface magnetic fields: The orange points show the rate at which net azithmuthally averaged subsurface toroidal magnetic flux in the northern hemisphere, $\langle b_\phi \rangle$, is generated by the poloidal field threading through the solar photosphere. For $\epsilon$~Eridani and 61~Cygni~A, error bars for the individual measuremnts were calculated using the closely spaced (in time) maps as discussed in the text. We comment that we get similar values of the errors for these two stars if we assume the difference between the fits and the measurements. 
The green points show the rate at which the net toroidal axisymmetric field in the northern hemisphere is lost through the surface due to flux emergence.  The solid curve shows the fits, and the dashed line shows the long-period component of the fit for $\epsilon$~Eridani.
The top panel shows the results obtained using the relatively high resolution images of the Sun, the  second row the reduced resolution Sun ($\ell \le 3$), the third row 61~Cygni~A, and the bottom row $\epsilon$~Eridani.  }
  \label{fig:Bphi_terms}
\end{center}
\end{figure}

To evaluate the integral
for the Sun, we used the measured surface differential rotation given by 
\citep{1988SoPh..117..291U}. 
The observed flux balance is shown for the Sun at the resolution of WSO and at the reduced resolution achievable for other stars in the top two rows of Figure~\ref{fig:Bphi_terms}.

For 61~Cygni~A and $\epsilon$~Eridani,
we used the same expression for the surface latitudinal differential rotation,
$(\Omega(\theta)-\Omega(\pi/2))$, as we used for the Sun. 
This assumption is reasonable to within a factor of two based on the literature.
For example,
\cite{2006ApJ...648..607C, 2016ApJ...824..150G} determined the differential rotation of $\epsilon$~Eridani from MOST observations. 
They found that the surface differential rotation is very similar to that of the Sun. ZDI inversions also estimate the differential rotation for both $\epsilon$~Eridani and 61~Cygni~A and it is within a factor of two of that of the Sun.
More generally, \cite{2013A&A...560A...4R}, using Kepler data, show that the differential rotation is only weakly sensitive to both the rotation rate and stellar type for cool stars. 
Turning to theory and numerical simulations, the mean-field hydrodynamic analysis by
\cite{1999A&A...344..911K} reports a weak dependence of differential rotation  $(\Omega(\theta)-\Omega(\pi/2))$ on rotation for K5 dwarfs -- with a scaling of $\Omega^{\beta}$ with $\beta$ in the range of -0.05 to 0.2. This is only about a 25\% change in the differential rotation between the fastest and slowest rotations of the Sun and the two K dwarfs. 
The same paper predicts a difference of a factor of about 2 between the G2 and K5 star. The assumption that $\epsilon$-Eridani and 61~Cygni~A have the same differential rotation as the Sun might lead to an overestimation of the rate at  which toroidal flux is generated, by a factor of up to 2.
Numerical simulations show that a bifurcation of the differential rotation occurs near a Rossby number of 1.0. In particular, the simulations indicate that stars with a Rossby number of more than about 1.0 should have anti-solar differential rotation \citep{2014MNRAS.438L..76G}.
The Rossby number listed for the Sun in Table~\ref{tab:param-stars} is larger than 1, which would naively lead to the conclusion that the Sun should have anti-solar differential rotation. However, the Rossby numbers listed are ultimately based on  mixing length models of convection. Such models greatly overestimate the convective velocities at large scales \citep{2012PNAS..10911928H}.
For this reason, the values for the Rossby numbers presented in Table~\ref{tab:param-stars}
are probably large overestimates. Since we know the Sun has solar-like differential rotation and the Rossby numbers of 61~Cygni~A and $\epsilon$~Eridani are smaller, we conclude that all three stars are likely to have solar-like differential rotation. The simulations then indicate only a weak dependence of the differential rotation 
on the rotation rate \cite{2014MNRAS.438L..76G} (noting their figure~2 is for the relative differential rotation). 

The other assumptions we make are that the toroidal flux observed on 61~Cygni~A and $\epsilon$~Eridani  is also due to flux emergence
and that $u_\mathrm{em}$ is the same as on the Sun.
As background to this assumption we note that realistic simulations of 
the turbulent convective velocities yield values near the photosphere of about 2 km/s for a G2V star and about 1km/s for a K5 star \citep{2013A&A...558A..48B}.
Our assumption that $u_\mathrm{em}$ is the same for all three stars might thus overestimate the rate of flux loss by a factor of up to two for $\epsilon$~Eridani and 61~Cygni~A.  The observed flux balances are shown in Figure ~\ref{fig:Bphi_terms}, and the amount generated by poloidal flux threading through the surface, and lost through
flux emergence are in approximate balance for each star, although the absolute amounts are greater for 61~Cygni~A and $\epsilon$~Eridani.

This application of Stokes' theorem does not tell us  whether latitudinal or radial differential rotation
(or some turbulent effect)
is responsible for the creation of the toroidal flux. It also does not tell us where the toroidal flux is generated,  or what role the tachocline plays in the dynamos.
What we have shown in this section is that the surface magnetic field plays a crucial role in the dynamo, not only of the Sun \citep{2015Sci...347.1333C}, but also for 61~Cygni~A and $\epsilon$~Eridani.

\section{{Cycle length}}
The longer time sequence of the S-index shown in the left-hand panels of Figure~\ref{fig:S_vs_Bphi} show that the Sun and 61~Cygni~A have had one dominant cycle length throughout the period from the late 1960s onwards, albeit with the strength of activity varying from cycle to cycle. $\epsilon$~Eridani is however different in that the short 3-year cycle only becomes comparable in amplitude to the longer cycle during the last 10 years. Our results show that $\epsilon$~Eridani has two coincident magnetic cycles that directly correspond to the shorter $\sim$3 year and longer $\sim$13 year chromospheric cycles.  The most recent S-index values for $\epsilon$ Eridani show values that are lower than predicted and further observations will indicate if $\epsilon$ Eridani is entering a period of inactivity despite its current very active state.

\section{Conclusions}

We have demonstrated that 61~Cygni~A and $\epsilon$~Eridani have S-index activity cycles that correlate with $\langle B_\phi \rangle_{(\phi,\theta)}$, the emergence of the unsigned net axisymmetric component of the toroidal field.   We further find that 
the surface poloidal magnetic field plays a crucial role in generating the toroidal field for 61~Cygni~A and $\epsilon$~Eridani, in a similar way as in the Sun. In all three stars, there is a  balance between the generation of toroidal flux associated with the poloidal field threading through the stellar surfaces and the loss of magnetic flux associated with flux emergence.  We also show the first magnetic butterfly diagram for stars other than the Sun.

\bibliography{refs.bib}
\bibliographystyle{aa.bst} 

{\noindent\bf{Acknowledgements:}} We firstly thank the referee for their constructive comments that helped improve the clarity of the paper.  We also thank Martina Coffaro for sending us her S-index values of $\epsilon$ Eridani so that we could ensure that the S-index values used in this work are compatible with previously published values.    SVJ acknowledges the support of the DFG priority programme SPP 1992 “Exploring the Diversity of Extrasolar Planets (JE 701/5-1)”.  RHC's contribution to this work was supported in part by ERC Synergy Grant WHOLE SUN 810218.  RHC benefited from discussions at the ISSI team 'What determines the dynamo effectively of solar active regions?' SBS acknowledges funding from the Austrian Science Fund (FWF) under the Lise Meitner grant M 2829-N. MMJ acknowledges support from STFC Consolidated Grant ST/R000824/1.  VS acknowledges funding from the European Research Council (ERC) under the European Unions Horizon 2020 research and innovation programme (grant agreement No. 682393 AWESoMeStars).  AAV acknowledges funding from the European Research Council (ERC) under the European Union's Horizon 2020 research and innovation programme (grant agreement No 817540, ASTROFLOW). Based on observations obtained at the Telescope Bernard Lyot (USR5026) operated by the Observatoire Midi-Pyrénées, Université de Toulouse (Paul Sabatier), Centre National de la Recherche Scientifique of France.  Wilcox Solar Observatory (WSO) data used in this study were obtained via the web site http:// wso.stanford.edu, courtesy of J. T. Hoeksema.    This work uses data obtained with the TIGRE telescope, located at La Luz observatory, Mexico. TIGRE is a collaboration of the Hamburger Sternwarte, the Universities of Hamburg, Guanajuato and Liège.  The Mount Wilson Observatory HK Project was supported by both public and private funds through the Carnegie Observatories, the Mount Wilson Institute, and the Harvard-Smithsonian Center for Astrophysics starting in 1966 and continuing for over 36 years.  These data are the result of the dedicated work of O. Wilson, A. Vaughan, G. Preston, D. Duncan, S. Baliunas, and many others.\\

{\noindent\bf{Data availability:}} The spectropolarimetic data are available from the Polarbase data archive: http://polarbase.irap.omp.eu/.

\newpage
\onecolumn
\begin{appendix}
\section{Chromospheric activity indices}

In this section we describe the sources for the time series of the large-scale magnetic field geometry and the chromospheric activity indicators of the Sun, 61~Cygni~A, and $\epsilon$~Eridani.  
For each of the three stars, we compiled a time series of measurements of the S-index, which is a measure of the chromospheric emission in the Ca II H\&K lines relative to the surrounding continuum.  The S-index was computed using the following relation  \cite{Duncan1991ApJS...76..383D} 
\begin{equation}
{\rm S} = \alpha(H + K)/(V +R)
\end{equation}
where the values for $H$ and $K$ are fluxes at the line cores, using triangular passbands, and $V$ and $R$ are the nearby continuum regions.  The central wavelengths for the Ca II H\&K lines are located at 393.363\,nm and  396.847\,nm with a width of 1.09\,nm.  The continuum regions ($V$ and $R$) used are at wavelengths 389.1--391.1\,nm and 399.1--401.1\,nm.  For each instrument, the values are calibrated to the Mount Wilson S-index values.   A list of the instruments used for the S-index time series and the dates of observations are summarised in Table~\ref{tab:mag-activity}. 

\FloatBarrier
\begin{table}[H]
\center
\caption{\label{tab:mag-activity} {\bf {Summary of S-index values used in this work.} }}
\begin{tabular}{llccccc}
\hline \hline
Star & Survey & Obs beg & Obs end & Span [days] & Nobs & Ref \\ 
\hline
Sun & MWO &  1966-09-01 & 2002-11-23  & 13231 & 237 & \cite{2017ApJ...835...25E}\\
 & NSO & 1976-11-20 & 2015-09-30 & 14193 & 4112 & \cite{2017ApJ...835...25E}\\
 & SSS & 1994-03-02 & 2016-02-26 & 8030  & 2000 & \cite{2017ApJ...835...25E}\\
61~Cygni~A & MWO & 1967-06-26 & 1994-11-14 & 10003 & 1122 &  {\cite{Baliunas1995ApJ...438..269B}}
\\
 & NARVAL & 2006-08-04 & 2018-07-22 & 4370  & 121
 &  {\cite{BoroSaikia2016A&A...594A..29B}}\\
 $\epsilon$~Eridani & MWO & 1967-18-10 & 2012-08-24 & 16350 & 41 & {\cite{Metcalfe2013ApJ...763L..26M}}  \\
 & SMARTS & 2007-08-22 & 2012-10-13 & 1880 & 141 & {\cite{Metcalfe2013ApJ...763L..26M}} \\
 & TIGRE & 2013-08-14 & 2019-03-02 & 2027 & 95 &
{\cite{Coffaro2020A&A...636A..49C}} \\
 & NARVAL & 2007-01-21 & 2018-11-16 & 4318 & 195 & this work \\
\hline
\hline
\end{tabular}
\end{table}
\tablefoot{MWO refers to observations secured at the Mount Wilson Observatory, NSO at the National Solar Observatory, SSS the Solar-Stellar Spectrograph.  The publications that contain the S-index values are shown in the column Ref.}

\end{appendix}

\end{document}